%% file: ajust.tex
\documentclass[11pt,a4paper]{article}

\usepackage{graphicx}
\usepackage[latin1]{inputenc}
\usepackage{latexsym}

\title{
\textbf{\large
Identification of dipolar relaxations in dielectric spectra of mid--voltage cross--linked polyethylene cables
}}

\author{
J. Òrrit,
J.C. Cañadas,
J. Sellarès %\thanks{e-mail: {\tt a.c.author@university.edu}, to whom all correspondence should be addressed.}
and J. Belana\\
\em\small{Departament de Física i Enginyeria Nuclear, Universitat Politècnica de Catalunya,}\\
\em\small{Campus de Terrassa, c. Colom 11, E-08222 Terrassa, Spain.}\\[0.25cm]
}

\date{}

\begin{document}

\maketitle

\begin{abstract}

Medium-voltage cross-linked polyethylene (MV-XLPE) cables have an important role in the electrical power distribution system. For this reason, the study of XLPE insulation is crucial to improve cable features and lifetime. Although a relaxational analysis using Thermally Stimulated Depolarization Currents (TSDC) can yield a lot of information about XLPE properties, sometimes its results are difficult to interpret. In previous works it was found that the TSDC spectrum of cables is dominated by a broad heteropolar peak, that appears just before an homopolar inversion, but the analysis of the cause of the peak was not conclusive. We have used a combination of TSDC and Isothermal Depolarization Currents (IDC) techniques to investigate further this issue. In order to discard spurious effects from the semiconductor interfaces, samples have been prepared in certain configurations and preliminary measurements have been done. Then, TSDC experiments have been performed using conventional polarization between 140~$^\circ$C and 40~$^\circ$C. Also, IDC measurements have been carried out between 90~$^\circ$C and 110~$^\circ$C in 2~$^\circ$C steps. The TSDC spectra show the broad peak at 95~$^\circ$C. On the other hand, IDC show a combination of power and exponential charge currents. Exponential currents are fitted to a Kohlrausch--Williams--Watts (KWW) model. The parameters obtained present approximately an Arrhenius behavior with $E_a=1.32$~eV, $\tau_0=3.29 \times 10^{-16}$~s, with a KWW parameter $\beta=0.8$. The depolarization current calculated from the obtained parameters turns out to match the dominant peak of TSDC spectra rather well. From the results and given the partially molten state of the material, we conclude that the most likely cause of the exponential IDC and the main TSDC peak is the relaxation of molecular dipoles from additives incorporated during the manufacturing process.

\end{abstract}


\newpage

\section{Introduction}
\label{intro}

Mid--voltage cross--linked polyethylene (MV--XLPE) cables are used in electricity distribution to connect electrical substations with primary customers or with transformers that serve secondary customers. Its use is widely spread since most of the electrical grid operates at this voltage range.

Although nowadays MV--XLPE cables are regarded as a commodity, some important points that affect their lifetime are not yet fully understood. Buildup of space charge in the bulk of the cable is known to be a major cause of failure \cite{failure}. For this reason, the study of space charge in MV--XLPE cables may provide clues about the most suitable composition (reticulants, antioxidants, \ldots) or manufacturing process (vulcanization time, \ldots) in order to enhance its operational lifetime \cite{marcador}. Relaxation, especially space charge relaxation, is a very relevant subject \cite{scr}, not only because of the role played by relaxational processes by themselves but also for the great amount of information about the electrical properties of the material that can be obtained from a relaxational study.

Among the techniques employed to study relaxations, thermally stimulated depolarization currents (TSDC) \cite{tsdc} stands out because of its high resolution and low equivalent frequency \cite{resolution}. A diagram that shows how a TSDC experiment is performed is presented in figure~\ref{figtsdc}. A TSDC experiment begins with a combination of electrical poling and thermal treatment so one or more relaxational mechanisms are activated in a metastable way \cite{wp}. Then, the sample is heated at a constant rate to force the relaxation of these mechanisms. Its relaxation gives rise to a displacement current that is recorded together with the temperature of the sample. This thermogram is the so--called TSDC spectrum and reflects the relaxations undergone by the sample.

TSDC has been employed successfully to find consistent differences in TSDC spectra of cable samples manufactured in the same factory under different conditions \cite{gc}.

TSDC is particularly useful to study overlapping relaxations since its high resolution allows to distinguish between relaxations with close relaxation times. This is what happens in the case of XLPE, where a large number of relaxations can be identified in the TSDC spectrum \cite{gc}. Unfortunately, some problems arise during the study of XLPE cables by TSDC that make it difficult to analyze the resulting spectra. 

In first place, TSDC spectra are strongly dependent on thermal history. This is a well--known characteristic of XLPE cables \cite{gc}. The effects of thermal history can be minimized performing each experiment on its own as--received sample.

Another problem is that cables are unusually thick compared with more commonly used samples. Since there is practical limit on the applied voltage that the experimental setup can supply, only relatively low electric fields can be employed to pole them. The space--charge or dipolar character of a relaxation can be determined studying how the polarization of the sample is related to the poling field \cite{dipolar} but in the low electric field limit space charge and dipolar relaxations behave in a similar way. As a consequence, this kind of analysis is not able to give conclusive results in this case. 

A way to overcome this difficulty is to employ a complementary technique to obtain information that can lead to the identification of one or several peaks. In this work we consider isothermal depolarization currents (IDC) as a complementary technique. There are several reasons that make this approach compelling. In first place, both techniques measure essentially the same physical effect, displacement current, so correlation between data obtained using both techniques is straightforward \cite{pet}.

Moreover, contributions from dipoles or space charge are much easier to distinguish in IDC data than in a TSDC spectrum \cite{taula}, especially if they are presented in a log--log plot. In theory, we can distinguish between three types of current in IDC data, that are presented in figure~\ref{figcurrents}. In section~\ref{generaleq} we will discuss the physical causes of each one of these currents, give the expressions that they follow and discuss its range of validity.

Of course, in practice it could be difficult to distinguish between these types of current since a meaningful log--log plot may require much more decades of data than the available ones.

Combining TSDC and IDC we expect to confirm previous assumptions about relaxations found in the TSDC spectrum of MV--XLPE cables \cite{gc} and to demonstrate how the correlation of TSDC with IDC data can improve our understanding of relaxation mechanisms in situations where the usual analysis of TSDC data is difficult, or even impossible, to carry out.

\section{Experimental}
\label{exp}

Cable was provided by General Cable from a single cable reel manufactured industrially. In this way, it was assured that the composition and the manufacturing process was the same for all the samples. The as--received cable consists, from inside to outside, of: (a) a $15.0$~mm of diameter aluminum core made up of nineteen twisted hexagonal wires, (b) an inner $0.50$~mm thick semiconducting layer made of a blend of ethylene-vinyl acetate (EVA), polyethylene (PE) and carbon black (CB), (c) a $15.5$~mm inner diameter and $24.6$~mm outer diameter insulating XLPE layer and (d) an outer $0.50$~mm thick semiconducting layer made of a blend of EVA, acrylonitrile and CB. The XLPE of the insulating layer was crosslinked mixing low density PE with a crosslinking agent (di--tert--butyl peroxide) and heating it up to a temperature higher than 200~$^\circ$C to produce a vulcanization reaction. Also, commercial additives, like antioxidants, are known to be present.

Figure~\ref{figdsc} presents a differential scanning calorimetry (DSC) scan of the XLPE insulation that shows that it begins to melt between 60~$^\circ$C and 70~$^\circ$C. The melting process has its maximum between 105~$^\circ$C and 110~$^\circ$C. In spite of this fact, the material remains consistent through this temperature range as a consequence of reticulation.

To make the sample, the cable was cut into 20~cm long sections and, then, a 2~cm long section of insulating and semiconducting layers was removed from one end. The semiconducting layers were used as electrodes. To avoid short circuits, the external layer was partially removed from the ends of the samples, leaving an 8~cm wide semiconducting strip centered in the sample. The cable core was used to make contact with the inner semiconducting layer. The contact with the outer semiconducting layer was made with an adjustable metallic clamp. A diagram showing the structure of the sample and the connection with the electrodes is presented in figure~\ref{samples}. The average mass of cable samples was $135$~g. The capacity of cable samples at $5$~kHz and $25$~$^\circ$C, measured with an HP 4192A LF impedance analyzer, was $27$~pF.

The experimental setups follow the diagram represented in figure~\ref{setups}. The IDC setup consists of a Köttermann 2715 forced air oven controlled by a Eurotherm 818P PID temperature programmer. Inside, two identical cable samples were placed close together in parallel position using a ceramic stand. One of them is used to host the temperature probe inside the insulating layer and measurements are performed on the other one. High voltage is supplied by a Brandenburg 807R (3--30~kV) potential source and the current is measured through a Keithley 6514 electrometer. The positive terminal of the source is connected to the cable core. The TSDC setup that was employed only differs in that it uses an Heraeus forced air oven and a Keithley 616 electrometer. 

In fact, both setups can be used either for IDC or TSDC. They are operated as follows. Switch~1 is connected to terminal~A during the poling stage and to terminal~B whenever the poling field is off. For IDC experiments, switch~2 is always connected to B so the charging of the sample can be monitored. For TSDC measurements, instead, switch~2 is operated in the same way as switch~1, to protect the electrometer.

IDC measurements were performed at several temperatures ($T_p$) between 90 and 110~$^\circ$C in 2~$^\circ$C steps. The poling time ($t_p$) was 1800~s and the discharge current was recorded for another 1800~s.
 
%
% Comentari thermal expansion
%
The effects of thermal expansion at the different temperatures does not need to be taken into account because it has been checked that the difference between the external diameter of the XLPE insulation at 90~$^\circ$C (24.8~mm) and 110~$^\circ$C (25.0~mm) is about 1\%. Once cooled, the external diameter does not return to its original value but to 24.6~mm, in both cases.
%
% Comentari paràmetres bàsics (massa/capacitat)
%
It has also been checked that due to annealing the change in mass is less than 1~g and the change in capacity ($5$~kHz, $25$~$^\circ$C) less than $1$~$pF$, in both cases

TSDC spectra were obtained by cooling down from 140~$^\circ$C ($T_p$) with a 12~kV applied voltage until 40~$^\circ$C ($\Delta T = 100$~$^\circ$C), and, after 5~min ($t_d$), heating up to 140~$^\circ$C again. Samples were poled without an isothermal poling stage so $t_a=t_p=0$. The heating or cooling rate of all the ramps was 2~$^\circ$C/min.

%
% Comentari condicions de mesura (temperatura)
%
The measurement conditions remained fairly constant throughout the experiments, with a room temperature around $25$~$^\circ$C.

Figure~\ref{figexemplecorba} presents an example of IDC experimental data. The best power law \cite{taula} fit is also plotted. The fit was very unsatisfactory and therefore it was necessary to consider other possibilities. Adding an exponential term to the power term was revealed as the best way to overcome this problem, as seen in figure~\ref{figcomponents}. The details will be given in the next section. However, to ensure that the effect that was measured had its origin on the dielectric properties of the cable insulation several previous experiments were performed.

An experiment was performed with a guard ring to ensure that the effect was not due to a superficial current. The exponential current was not affected.

Next, IDC measurements were performed on $2 \times 2$~cm samples with thickness 160~$\mu$m, at 92~$^\circ$C and poling with a 1~kV electric potential. Samples were obtained using a lathe with a special cut tool to convert the XLPE insulating layer into a roll and circular aluminum electrodes of 1~cm of diameter were vacuum deposited at the center of both sides of the samples. The data obtained did not show exponential behavior. Then, the question we had to answer was which one of these three elements was the cause of the exponential current: size, cylindrical shape or electrodes (semiconductors). 

To address this question, we performed IDC measurements at 92~$^\circ$C with a 10~kV electric potential, in first place with a cable from which we had removed the external semiconducting layer. Then, we did the same with another sample without the inner one, removing the core by traction and the semiconducting layer with a drill. Finally, we removed both semiconducting layers from a cable sample and we repeated the same experiment. The curves that were recorded show the exponential relaxation in all three cases.  

Once we discarded the influence of any semiconductor layer we had to consider the effect of cylindrical shape. To this end, we cutted from the XLPE insulation a cuboid shaped sample of $4.5 \times 4.5 \times 50$~mm and vacuum deposited two electrodes on opposite faces of $4.5 \times 50$~mm. By performing measurements in the same conditions than in the previous tests, we found the existence of the exponential relaxation again. Consequently, the size turned out to be the determining factor for the appearance of such phenomenon. This was finally corroborated by performing a measurement with a  $2 \times 2$~cm sample with thickness 2~mm. Unlike the response of 16~$\mu$m thick samples, in this case the exponential relaxation was detected ---although not very pronounced---. In these experiences a field of $2.2$~kV/mm was applied by modulating the voltage proportionally to the thickness.

\section{Results and discussion}
\label{resdis}

% We read and understood the papers \cite{insulators, transport, hnkww}.

\subsection{General equations}
\label{generaleq}

As a first approximation, we can assume that polarization is a first--order process, described by
\begin{equation}
\frac{dP}{dt} = \frac{P_{eq}-P}{\tau}.
\label{pedete}
\end{equation}
In the case of depolarization $P_{eq} = 0$ and equation~\ref{pedete} becomes
\begin{equation}
\frac{dP}{dt} = - \frac{P}{\tau}.
\label{depedete}
\end{equation}
The solution of this equation can be written as
\begin{equation}
P(t) = P_0 \phi[z(t)],
\label{soldepedete}
\end{equation}
in terms of the {\em reduced time}
\begin{equation}
z(t) \equiv \int_0^t \frac{dt}{\tau}
\label{reducedtime}
\end{equation}
and the {\em dielectric decay function}
\begin{equation}
\phi(x) \equiv \exp(-x)
\label{firstorderdecayfunction}
\end{equation}
The {\em displacement current density} is defined as
\begin{equation}
J = - \frac{dP}{dt}.
\label{currentdefinition}
\end{equation}
It can also be written in terms of the reduced time and the dielectric decay function, as
\begin{equation}
J = - P_0 \phi'(z) \frac{dz}{dt} = - P_0 \frac{\phi'(z)}{\tau} ,
\label{currentgeneralexpression}
\end{equation}
where in the last equality we have substituted the derivative of $z(t)$ by its expression.
From equation~\ref{firstorderdecayfunction} we have that
\begin{equation}
\phi'(x) = - \exp(-x)
\label{firstorderdecayfunctionderivative}
\end{equation}
and therefore
\begin{equation}
J = P_0 \frac{\exp(-z)}{\tau} ,
\label{firstordercurrent}
\end{equation}
which is the first--order displacement current density. 

The previous equations describe an exponential relaxation. They can be modified to describe the so--called stretched exponential behavior represented by the Kohlrausch--Williams--Watts (KWW) model. Within this model, equation~\ref{firstordercurrent} is generalized replacing the dielectric decay function given by equation~\ref{firstorderdecayfunction} by a more general expression
\begin{equation}
\phi_\beta(x) \equiv \exp(-x^\beta)
\label{kwwdecayfunction}
\end{equation}
where $\beta>0$. Although in its origin KWW was introduced as an empirical improvement to existing models, nowadays it is interpreted as a way to take into account a distribution of relaxation times \cite{hnkww}. In the case $\beta=1$ we fall back into the proper exponential relaxation described by equation~\ref{firstorderdecayfunction}. We will refer to any of these relaxations as exponential because the shape of the current curve is similar but we will use the KWW model in all the fits since it provides a more accurate description of the current.

The derivative of the dielectric decay function becomes
\begin{equation}
\phi'_\beta(x) \equiv - \beta x^{\beta - 1} \exp(-x^\beta)
\label{kwwdecayfunctionderivative}
\end{equation}
and, as a consequence, the KWW displacement current density is
\begin{equation}
J = P_0 \beta z^{\beta - 1} \frac{ \exp(-z^\beta)}{\tau} .
\label{kwwcurrent}
\end{equation}

We can now apply equation~\ref{kwwcurrent} to two interesting cases: isothermal depolarization currents (IDC) and thermally stimulated depolarization currents (TSDC). In the first case we can safely assume that $\tau$ does not change with time, since this is the behavior predicted by the Arrhenius or the WLF models. This leads to a great simplification of equation~\ref{reducedtime} for the IDC case
\begin{equation}
z(t) = \frac{t}{\tau}.
\label{idcreducedtime}
\end{equation}
Substituting in equation~\ref{kwwcurrent} we obtain
\begin{equation}
J(t) = P_0 \beta \frac{t^{\beta - 1}}{\tau^\beta} \exp\left[ - \left( \frac{t}{\tau} \right)^\beta \right]
\label{idckwwcurrent}
\end{equation}

This is the stretched exponential current, or, simply, exponential current. In a log--log plot it has only a horizontal asymptote for short times.

In fact, we can expect that the IDC has also a free charge component that we denote as power current \cite{taula, isotermes}
\begin{equation}
J(t) = C \, t^\alpha \left[ 1 - \left( \frac{1+t_p}{t} \right)^\alpha \right]
\end{equation}
where $\alpha <0$. In a log--log plot it appears with two asymptotes. For short times it approximates to an oblique asymptote while for long times it tends to a vertical asymptote.

Therefore we will fit the IDC to
\begin{equation}
J(t) = C \, t^\alpha \left[ 1 - \left( \frac{1+t_p}{t} \right)^\alpha \right] + D \, t^{\beta - 1} \exp\left[ - \left( \frac{t}{\tau} \right)^\beta \right].
\end{equation}
Since intensity is proportional to density current, through the area of the electrode, we can fit intensity curves using this expression.

The second case is somewhat more complicated because depolarization takes place during a heating ramp. Assuming a constant heating rate $v$, temperature will be given by
\begin{equation}
T(t) = T_d + v t.
\label{heatingramp}
\end{equation}
Since it is usual to plot $J$ in terms of $T$, we express $z$ as a function of temperature
\begin{equation}
z(T) = \frac{1}{v} \int_{T_d}^T \frac{dT}{\tau(T)}
\label{tsdcreducedtime}
\end{equation}
so we can compare the experimental plot to \cite{amorphous}
\begin{equation}
J(T) = P_0 \exp \left\{ - z^\beta(T) + \ln \left[ \frac{\beta}{\tau(T)} z^{\beta -1}(T) \right] \right\}.
\label{tsdckwwcurrent}
\end{equation}
As in the previous case $I(T) \propto J(T)$ and the expression can be applied to intensity curves.

Among the possible causes of a power law discharge current are dipole depolarization, Maxwell--Wagner relaxation, electrode polarization or recombination of trapped space charge \cite{isotermes}. 

Exponential current is often due to depolarization of molecular dipoles but it can also be due to recombination of trapped space charge, that is taken out of its traps by thermal excitation, when there is no probability of retrapping \cite{randall}. In fact, recombination without retrapping happens when the displacement of space charge during polarization has been so small that trapped charges act like small dipoles, sometimes called Gerson dipoles \cite{gerson}. Therefore we will attribute exponential current to dipolar relaxation, including under this denomination depolarization of molecular dipoles and recombination of Gerson dipoles.

When there is a strong probability of retrapping, the recombination of trapped space charge must be modeled as a higher--order process \cite{garlick}. We can obtain the IDC current from the depolarization, given by
\begin{equation}
\frac{dP}{dt} = - \frac{P^n}{\tau}
\label{depedeteordren}
\end{equation}
where $\tau$ can no longer be interpreted as a relaxation time. The solution of this equation assuming that $\tau$ is constant (isothermal depolarization) is
\begin{equation}
P(t) = \left[ P_0^{1-n} - (1-n) \frac{t}{\tau} \right]^{\frac{1}{1-n}}
\label{polarizationinversesquare}
\end{equation}
and
\begin{equation}
J(t) = \left[ J_0^{\frac{1-n}{n}} - (1-n) \frac{t}{\tau^{1/n}} \right]^{\frac{n}{1-n}}
\label{densitycurrentinversesquare}
\end{equation}
A current arising from trapped space charge recombination with a high retrapping probability, would correspond to the case $n=2$
\begin{equation}
J(t) = \left[ J_0^{-\frac{1}{2}} + \frac{t}{\tau^{1/2}} \right]^{-2}.
\end{equation}
We will refer to such a current as inverse square current. Its log--log plot tends to a horizontal asymptote for short times and to an oblique asymptote for long times. In fact, this kind of current is often modeled better using an empirical value for $n$ between $1$ and $2$, that would indicate an intermediate case between no retrapping and strong retrapping probability. Such empirical value would replace the $2$ in the exponent for a higher integer. Nevertheless, the shape of the current would be the same so be will refer anyway to any current of this kind as inverse square current.

It should be emphasized that the inverse--square and the exponential cases imply a narrow distribution of relaxation times. When there is a broad distribution of relaxation times, either space charge or dipoles can yield a power current.

\subsection{Data analysis}

In figure~\ref{figmostraajustos} five experimental curves are presented, together with the result of their fit. These curves represent well the behavior shown by the eleven isothermal experiments performed between $90$~$^\circ$C and $110$~$^\circ$C.

These temperature limits have been chosen because under $90$~$^\circ$C the exponential relaxation is hardly noticeable. On the other hand, above $110$~$^\circ$C homopolar currents appear making it very difficult to analyze the experiments. More or less, this temperature range coincides with the fusion peak studied by DSC, that begins at $90$~$^\circ$C and has a maximum at $110$~$^\circ$C. Incidentally, this temperature range also includes the operating temperature of the cables.

It is very difficult to establish the exact nature of the power current. In fact, it could also be an inverse square current since it is not easy to register enough decades of data to distinguish between both types of current. A plausible explanation is that it is due to electrode polarization. It is very hard to find electrodes that are completely transparent to current. Therefore a certain amount of charge is probably retained at the electrodes. Its depolarization can give rise to the observed power current. Recombination of trapped space charge is also possible, especially if it is due to disappearance of traps due to the melting of the material.

Nevertheless, the task of determining the exact nature of the process that yields this current is a very difficult one \cite{transport}. Moreover, it does not matter for the end of this work, because we are focused only on currents that can cause the 95~$^\circ$C peak, and this is not the case for the power current.

The exponential relaxation shows the typical behavior of a thermally activated process. For lower temperatures it shows up later. As a consequence, at low temperatures the power current determines the IDC response for short times while the exponential current is the responsible of the larger time response. Instead, at higher temperatures the opposite interplay between currents occurs, as it can be seen in figure~\ref{figmostraajustos}.

The numeric result of all the fits is presented in table~\ref{table1}. The values of $C$ tend to diminish with increasing temperatures while $D$ remains more or less constant. The parameter $\alpha$ that characterizes the power current changes as a consequence of structural change, most probably the fusion of the material. Instead, the parameter $\beta$ does not change significantly throughout all the experiments. This can be interpreted as a sign that the exponential current is not related to XLPE itself but to some other component incorporated at the manufacturing process. Anyway, we will assume that $\beta$ does not depend on temperature and we will take it as a characteristic parameter of the relaxation.

We can assume that the behavior of the $\tau$ parameter in table~\ref{table1} follows Arrhenius law
\begin{equation}
\tau = \tau_0 \exp \left( \frac{E_a}{k T} \right)
\end{equation}
as it can be seen in figure~\ref{figarrhenius}. The linear regression plotted in this figure reads $\ln(\tau) = -35.7 + 1.53 \times 10^4 / T$, this is, $\tau_0 = 3.29 \times 10^{-16}$~s and $E_a = 1.32$~eV.

\subsection{Discussion}

We have seen that, aside from the usual power current, IDC experiments show an exponential current that can be fitted successfully to a KWW model. Through these fits, a relaxation time for the exponential current can be obtained for each IDC experiment. The KWW parameter itself seems to be constant, allowing a great simplification of the data analysis. As a consequence, the exponential current can be described in terms of three parameters: $E_a$, $\tau_0$ and $\beta$.

Equation~\ref{tsdckwwcurrent} can be employed to predict the shape of the TSDC current in terms of the obtained parameters. Of course, we will obtain just one of the many peaks present in the TSDC spectrum but if we compare the calculated current with the experimental spectrum we can identify the TSDC peak that corresponds to the exponential current found by IDC.

This comparison can be seen in figure~\ref{figcomparacio}. We can see that the predicted KWW peak fits rather well to the main peak of the TSDC spectra, placed at $368$~K. A $3$~K shift towards lower temperatures has been applied to the predicted curve to obtain a better concordance. This difference can be due to a temperature gradient inside the oven or, simply, to uncertainty in the fit results. As usual in most relaxation models, the KWW model tends to underestimate the amount of current before the maximum of the peak. Other than these two disparities, the agreement between both peaks is noticeable.

Taking this into account, we state that the IDC exponential current and the 95~$^\circ$C TSDC peak are due to the same physical cause. We will discuss in the following lines which could be that cause.

The most probable cause of the exponential current is polarization of molecular dipoles in the cable bulk. We have seen that usually dipolar currents adopt an exponential form, whenever the distribution of relaxation times has a narrow shape. 

The shape of the current is also compatible with recombination of Gerson dipoles but since the material is partially molten it does not seem feasible that there exists an stable trap structure that could give a well--behaved exponential current. More probably, any trapped space--charge would give rise to a more irregular peak \cite{recryst}.

Nevertheless, two questions arise from this explanation. In first place, why does the relaxation show on full cables but not on thin samples obtained from the cables themselves? In second place, if XLPE is a non polar material, where do the dipoles come from?

The first question suggests that the intensity of the power current is proportional to the surface. Since the intensity of the exponential current should be proportional to the volume, if it comes from dipolar relaxation, that would explain why it is not visible on samples with a high surface--to--volume ratio. As explained previously, it is not our goal to dilucidate the origin of the power current but we will point out that proportionality to the surface is to be expected if it comes from electrode polarization, just to mention one possible cause.

The second question can be explained if we assume that the exponential current comes from additives present in the bulk of the cable, either introduced during the manufacturing process (traces of reticulant, antioxidant, \ldots).

To confirm this hypothesis, several TSDC experiments have been performed on the same cable sample. The spectra is plotted in figure~\ref{figpassades}. It can be seen that after the first experiment the peak at 95~$^\circ$C disappears. Even though the subsequent spectra keep changing, the peak does not show up again. The most likely cause of the peak is, therefore, some additive or by--product that does not stand a single heating ramp until 140~$^\circ$C. 

On the other hand, we discard from previous experiments described in section~\ref{exp} that the current comes from the semiconductor layers. Moreover, since these layers are conductive to a certain degree it is not probable that a large electric field builds inside these layers during polarization.

\section{Conclusions}

In a previous work about the TSDC spectrum of MV XLPE cables, it was found that the large heteropolar peak placed before the homopolar peak presented a dipole--like behavior \cite{gc}. The main evidence was that polarization turned out to be proportional to the applied field, as it usually happens with dipoles. Anyway, space charge polarization also tends to be proportional to the applied field for low field values.

For this reason it was interesting to seek confirmation for these findings using another technique. With IDC, an exponential current has been found that corresponds to the aforementioned peak. Thus, the dipolar character of the peak is confirmed by a complementary technique.

Unfortunately, IDC does not give further clues on the exact nature of the exponential current. In fact, the dipolar current could be produced either by molecular dipoles or dipoles created by microscopic displacement of space charge. Anyway, the stability of the current in a wide temperature range, the relatively low signal obtained and the way the peak depends on thermal history allows us to infer that its cause is not in the XLPE itself but in some additive introduced during the manufacturing process. This would also imply that the relaxation is due to molecular dipoles since it is unlikely that the additive has charge trapping capabilities.

In spite of this fact, correlation of IDC data with TSDC data shows promise in the study of systems where high electric fields can not be applied due to the large dimensions of the sample. 

Finally, we will stress that a proper characterization of the relaxations of MV XLPE cables is instrumental in industrial applications such as monitoring of cable degradation. Further research on this subject could lead to interpretation of the spectra that would allow to assess the state of cables in a non--destructive way \cite{marcador}.\\[0.25cm]

{\bf Acknowledgments} This work has been partially supported by project {\em 2009 SGR 01168} ({\em AGAUR}, Generalitat de Catalunya). The authors thank General Cable for providing cable samples.

\bibliography{ajust}

\newpage

\pagestyle{empty}

\begin{table}[h]

\caption{Fit results of the IDC curves. \label{table1}}

\begin{center}

\begin{tabular}{ccccccc}

$T$ & $C$~(A) & $\alpha$ & $\beta$ & $D$~(A) & $\tau$~(s) \\
90  & $1.01 \times 10^{-10}$ & $-1.29$ & $0.82$ & $8.66 \times 10^{-11}$ & $7.59 \times 10^{2}$ \\
92  & $4.12 \times 10^{-11}$ & $-0.09$ & $0.77$ & $2.02 \times 10^{-10}$ & $5.29 \times 10^{2}$ \\
94  & $1.56 \times 10^{-11}$ & $-0.15$ & $0.84$ & $1.71 \times 10^{-10}$ & $3.37 \times 10^{2}$ \\
96  & $2.35 \times 10^{-11}$ & $-0.12$ & $0.77$ & $1.95 \times 10^{-10}$ & $3.10 \times 10^{2}$ \\
98  & $6.33 \times 10^{-12}$ & $-0.13$ & $0.78$ & $2.11 \times 10^{-10}$ & $3.07 \times 10^{2}$ \\
100 & $1.43 \times 10^{-11}$ & $-0.30$ & $0.79$ & $2.78 \times 10^{-10}$ & $2.81 \times 10^{2}$ \\
102 & $8.98 \times 10^{-12}$ & $-0.14$ & $0.80$ & $2.87 \times 10^{-10}$ & $1.84 \times 10^{2}$ \\
104 & $6.41 \times 10^{-12}$ & $-0.30$ & $0.88$ & $7.90 \times 10^{-11}$ & $1.19 \times 10^{2}$ \\
106 & $2.46 \times 10^{-11}$ & $-0.41$ & $0.76$ & $2.12 \times 10^{-10}$ & $1.12 \times 10^{2}$ \\
108 & $3.20 \times 10^{-12}$ & $-0.48$ & $0.83$ & $2.90 \times 10^{-10}$ & $1.02 \times 10^{2}$ \\
110 & $3.56 \times 10^{-12}$ & $-0.46$ & $0.87$ & $2.55 \times 10^{-10}$ & $7.01 \times 10^{1}$ \\

\end{tabular}

\end{center}

\end{table}

\clearpage

\newtheorem{peudefigura}{Figure}

\begin{peudefigura}
{\rm Representation of a TSDC experiment: ($t_a$) annealing time; ($t_p$) polarization time; ($t_d$) deposit time; ($T_p$) polarization temperature; ($\Delta T$) polarization temperature range; ($T_d$) deposit temperature; ($E_p$) poling field.}
\label{figtsdc}
\end{peudefigura}

\begin{peudefigura}
{\rm Log-log plot of three types of currents: ($-$) exponential; ($--$) inverse square; ($\cdot - \cdot$) power.}
\label{figcurrents}
\end{peudefigura}

\begin{peudefigura}
{\rm DSC thermogram for a $30.15$~mg XLPE cable insulation sample showing the fusion peak at 108~$^\circ$C. The heating rate was 2~$^\circ$C/min.}
\label{figdsc}
\end{peudefigura}

\begin{peudefigura}
{\rm Diagram of the sample and of the connection with the electrodes: (a) cable core; (b) inner semiconducting layer; (c) XLPE insulating layer; (d) outer semiconducting layer; (e) adjustable clamp. The arrows represent the direction of the electric field at the zone where it is more intense.}
\label{samples}
\end{peudefigura}

\begin{peudefigura}
{\rm Diagram of the experimental setup: (HVS) high voltage source; (FAO) forced air oven; (S) sample; (PID) temperature programmer; (TP) temperature probe; (A) electrometer (connected to analog--to--digital port of computer).}
\label{setups}
\end{peudefigura}

\begin{peudefigura}
{\rm IDC curve for $T_p=104$~$^\circ$C, $t_p=1800$~s and $V_p=10$~kV in a log-log diagram: ($-$) experimental; ($--$) power law fit.}
\label{figexemplecorba}
\end{peudefigura}

\begin{peudefigura}
{\rm ($\circ$) IDC curve for $T_p=104$~$^\circ$C, $t_p=1800$~s and $V_p=10$~kV in a log--log diagram; ($-$) theoretical fit; ($--$) KWW component; ($\cdot - \cdot$) power law component.}
\label{figcomponents}
\end{peudefigura}

\begin{peudefigura}
{\rm IDC curves and theoretical fits in a log-log diagram for $t_p=1800$~s and $V_p=10$~kV and $T_p$: ($\circ$) 92 $^\circ$C; ($\Box$) 96 $^\circ$C; ($\Diamond$) 100 $^\circ$C; ($\triangle$) 104 $^\circ$C; ($\triangleleft$) 108 $^\circ$C.}
\label{figmostraajustos}
\end{peudefigura}

\begin{peudefigura}
{\rm Arrhenius plot: ($\circ$) relaxation time versus $T^{-1}$; ($-$) linear regression.}
\label{figarrhenius}
\end{peudefigura}

\begin{peudefigura}
{\rm ($-$) TSDC curve for $T_p=140$~$^\circ$C, $T_d=40$~$^\circ$C, $t_d=5$~min and $v=2$~$^\circ$C/min; ($--$) simulated TSDC from equation~\ref{tsdckwwcurrent} and for $E_a=1.32$~eV, $\tau_0=3.29 \times 10^{-16}$~s and $\beta=0.8$.}
\label{figcomparacio}
\end{peudefigura}

\begin{peudefigura}
{\rm TSDC curves for $T_p=140$~$^\circ$C, $T_d=40$~$^\circ$C, $t_d=5$~min and $v=2$~$^\circ$C/min. The experience number with the sample is given next to the curve.}
\label{figpassades}
\end{peudefigura}

\clearpage

\input{ajust-figures.tex}

\end{document}

%% file: ajust-figures.tex
\newcommand{\dibuix}[2]{%

\newpage

\pagestyle{empty}

\hbox{}\vspace{2cm}

\begin{center}
\includegraphics[width=10cm]{#1}
\end{center}

\vspace{3cm}

\noindent
{\bf Figure #2}

\noindent
J. Òrrit, J.C. Cañadas, J. Sellarès and J. Belana, ``Identification of dipolar relaxations in dielectric spectra of mid--voltage cross--linked polyethylene cables''.

}
\dibuix{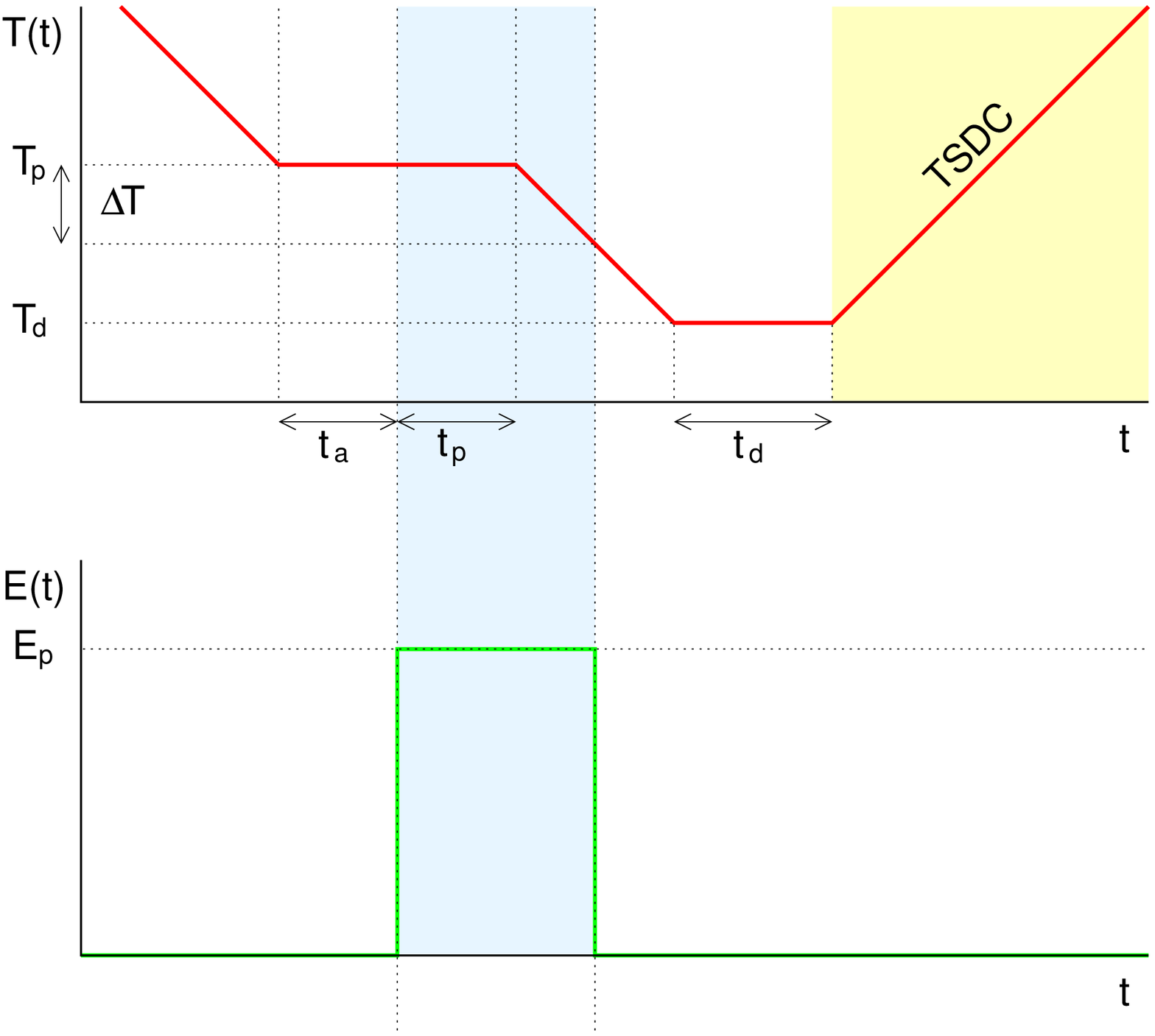}{\ref{figtsdc}}

\dibuix{currents}{\ref{figcurrents}}

\dibuix{dsc}{\ref{figdsc}}

\dibuix{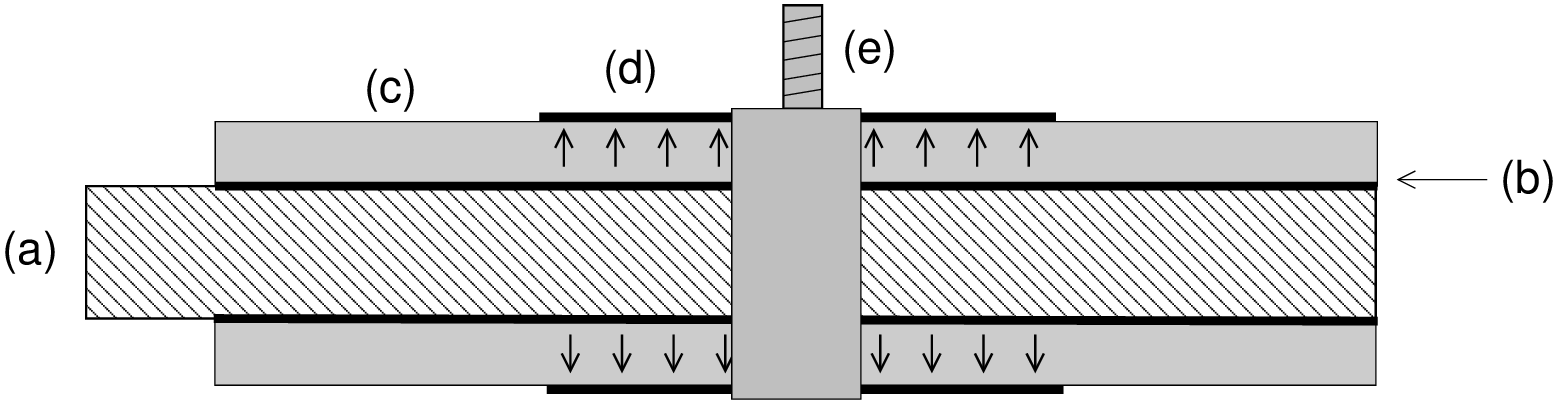}{\ref{samples}}

\dibuix{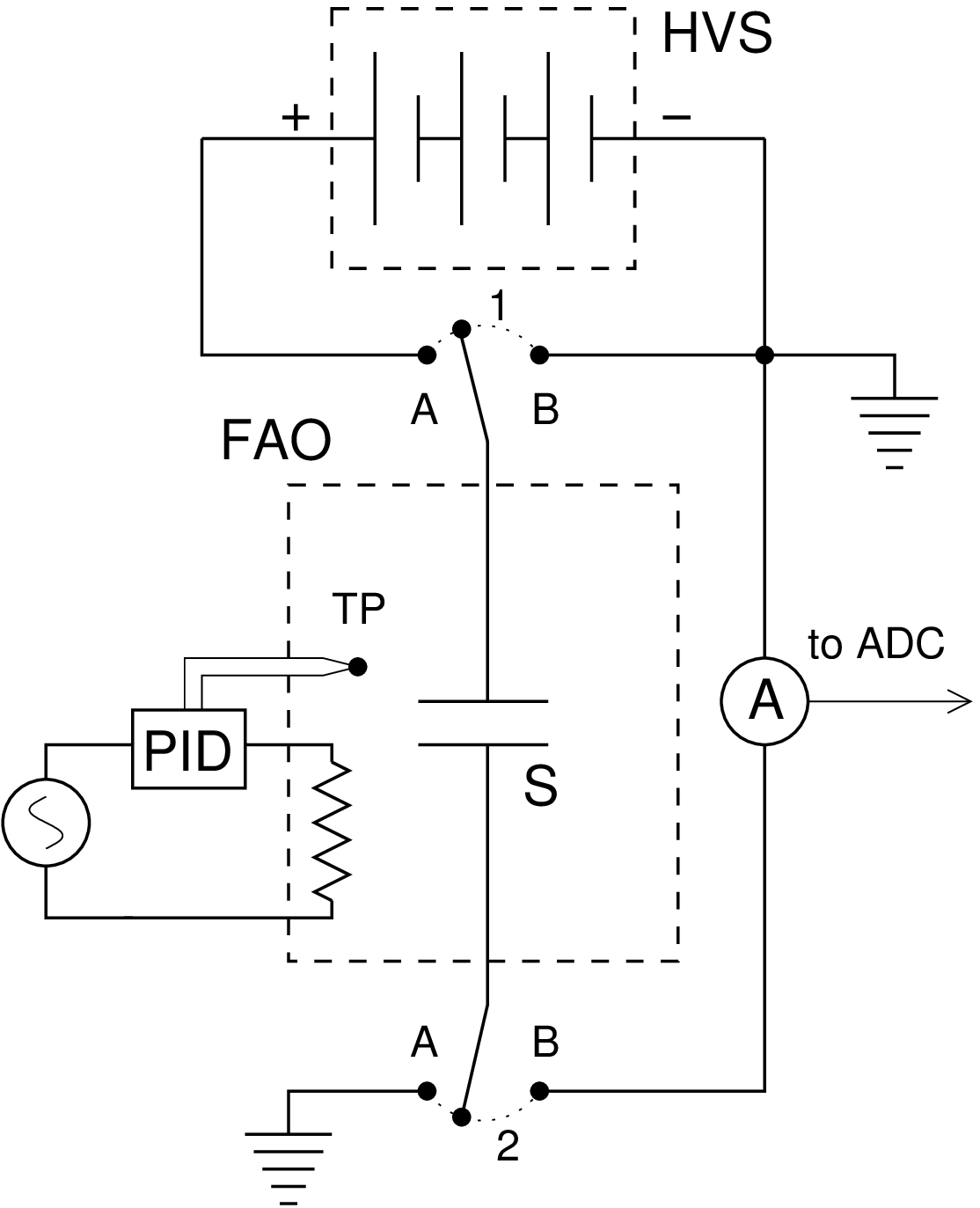}{\ref{setups}}

\dibuix{exemplecorba}{\ref{figexemplecorba}}

\dibuix{components}{\ref{figcomponents}}

\dibuix{mostra-ajustos}{\ref{figmostraajustos}}

\dibuix{arrhenius}{\ref{figarrhenius}}

\dibuix{comparacio}{\ref{figcomparacio}}

\dibuix{passades}{\ref{figpassades}}

%\dibuix{...}{...}